\pdfoutput=1
\documentclass[11pt,twoside,a4paper,cmspaper,final,collab]{cms-tdr}
\def\svnVersion{6d7f2ac}\def\svnDate{2024/06/25}\def\cmsCernNoTag{CERN-EP-2023-186}\def\cmsCernDate{\today}\def\cmsMessage{Published in Physical Review D as \href{http://dx.doi.org/10.1103/PhysRevD.109.112009}{\doi{10.1103/PhysRevD.109.112009}.}}
\begin{document}\cmsNoteHeader{SUS-19-001}

\newcommand{\nJets}{\ensuremath{N_{\text{jets}}}\xspace}
\newcommand{\NEvtsNorm}{\ensuremath{N^{\text{evts}}(\nJets,\allowbreak 1200<\st<1300\GeV)}\xspace}
\DeclareRobustCommand{\PS}{{\HepParticle{S}{}{}}\Xspace}
\DeclareRobustCommand{\PSS}{{\HepSusyParticle{\PS}{}{}}\Xspace}
\newcommand{\st}{\ensuremath{S_{\mathrm{T}}}\xspace}
\newcommand{\gluinoMass}{\ensuremath{m_{\sGlu}}\xspace}
\newcommand{\squarkMass}{\ensuremath{m_{\sQua}}\xspace}
\newcommand{\neutralinoMass}{\ensuremath{m_{\PSGczDo}}\xspace}
\newcommand{\fshape}{\ensuremath{f^{\mathrm{AGK}}(\st\,\text{bin}\,i)}}
\newcommand{\radjustment}{\ensuremath{r(\nJets,\st\,\text{bin}\,i)}\xspace}
\newcommand{\AGKTemplate}{\ensuremath{K(\st)}\xspace}

\ifthenelse{\boolean{cms@external}}{\providecommand{\cmsLeft}{upper\xspace}}{\providecommand{\cmsLeft}{left\xspace}}
\ifthenelse{\boolean{cms@external}}{\providecommand{\cmsRight}{lower\xspace}}{\providecommand{\cmsRight}{right\xspace}}

\cmsNoteHeader{SUS-19-001}

\title{Search for stealth supersymmetry in final states with two photons, jets, and low missing transverse momentum in proton-proton collisions at \texorpdfstring{$\sqrt{s}=13\TeV$}{sqrt(s) = 13 TeV}}

\date{\today}

\abstract{The results of a search for stealth supersymmetry in final states with two photons and jets, targeting a phase space region with low missing transverse momentum (\ptmiss), are reported. The study is based on a sample of proton-proton collisions at $\sqrt{s}=13\TeV$ collected by the CMS experiment, corresponding to an integrated luminosity of 138\fbinv. As LHC results continue to constrain the parameter space of the minimal supersymmetric standard model, the low \ptmiss~regime is increasingly valuable to explore. To estimate the backgrounds due to standard model processes in such events, we apply corrections derived from simulation to an estimate based on a control selection in data. The results are interpreted in the context of simplified stealth supersymmetry models with gluino and squark pair production. The observed data are consistent with the standard model predictions, and gluino (squark) masses of up to 2150 (1850)\GeV are excluded at the 95\% confidence level.}

\hypersetup{%
pdfauthor={CMS Collaboration},%
pdftitle={Search for stealth supersymmetry in final states with two photons, jets, and low missing transverse momentum in proton-proton collisions at sqrt(s)=13 TeV},%
pdfsubject={CMS},%
pdfkeywords={CMS, supersymmetry, SUSY, stealth, gluino, squark}}

\maketitle

\section{Introduction}
\label{sec:introduction}

The search for signatures of models featuring supersymmetry (SUSY) with a stable lightest SUSY particle (LSP)~\cite{Farrar:1978xj} is an important part of the physics program at the CERN LHC. Such models can solve the hierarchy problem, can allow for the unification of the gauge couplings, and can provide a candidate for dark matter. Typically, such signatures involve a significant imbalance in the momenta of detected particles in a collision, because LSPs do not interact with the material of the detector. Therefore, many searches for SUSY focus on events that feature a significant imbalance in the sum of the momenta in the transverse plane, called the missing transverse momentum \ptvecmiss, whose magnitude is defined as \ptmiss~\cite{Dimopoulos:1996gy, Martin:1996zb, Poppitz:1996xw, Meade:2008wd, Buican:2008ws, Abel:2009ve, Carpenter:2008wi, Dumitrescu:2010ha}.

On the other hand, there are many SUSY models that do not exhibit this signature, such as those with compressed spectra~\cite{LeCompte:2011cn, LeCompte:2011fh}, $R$-parity violation~\cite{Farrar:1978xj, Barbier:2004ez, Graham:2014vya}, hidden valleys~\cite{Strassler:2006im, Han:2007ae, Nakai:2015swg}, and others~\cite{Burdman:2006tz, Graham:2012th, Alves:2013wra, Ellwanger:2014hia, Heidenreich:2014jpa}. Given recent results imposing progressively higher bounds on gluino and squark masses (\gluinoMass~and \squarkMass) in models of SUSY with high-\ptmiss~signatures~\cite{CMS:2020bfa, CMS:2020fia, CMS:2020cpy, CMS:2020cur, CMS:2019zmd, CMS:2019ybf}, a thorough study of such alternatives becomes important. One such alternative is the stealth scenario~\cite{Fan:2011yu, Fan:2012jf, Fan:2015mxp}, in which the minimal supersymmetric standard model (MSSM) is augmented with a light hidden sector. This sector only couples weakly to the SUSY-breaking sector. Consequently, the particles in this sector are nearly mass degenerate with their SUSY partners. The simplest hidden sector consists of a single scalar boson, called a singlet \PS, and its corresponding SUSY fermion, called a singlino \PSS. As the mass degeneracy between them is broken only through suppressed or second-order diagrams, their mass difference $\Delta m(\PS,\PSS)$ is naturally expected to be small. Previously, searches have been performed for stealth SUSY in events with top quarks and several light-flavor partons~\cite{CMS:2021knz}. A previous study was also performed for stealth SUSY with jets, photons or leptons, and low \ptmiss at $\sqrt{s}=8\TeV$~\cite{CMS:2014exa}. In that analysis, the background model for the diphoton final state used an assumption called \st~scaling, which is described in Section~\ref{sec:backgroundEstimation}. The present study relaxes the assumption of \st~scaling and uses simulated background samples to estimate the corrections needed to a background model derived from this assumption for a diphoton final state.

In the stealth SUSY models considered in this study, the gravitino \PXXSG is the LSP, and is produced through the decay of the stealth sector singlino, $\PSS \to \PXXSG\PS$. Because the mass splitting $\Delta m(\PS,\PSS)$ is small compared to the masses of \PS~and \PSS, the momentum of the LSP is suppressed by the small kinematic phase space available. As \ptmiss arises predominantly from the unobserved LSPs in the final state, stealth SUSY decays will lead to lower average values of \ptmiss than their regular SUSY counterparts. This is true even when the singlino is Lorentz boosted by initial state radiation (ISR): in a two-body decay with small mass splitting between the parent and the heavier of its daughters, the light daughter (here the LSP) acquires a small fraction of the parent's energy. This is in contrast with a typical compressed-spectrum SUSY model in which the heavy daughter is the LSP.

Supersymmetric particles can decay through the hidden sector without violating $R$-parity, with an MSSM electroweakino---an MSSM partner of a standard model (SM) gauge boson---decaying to the \PSS via the emission of an SM gauge boson. In this paper, we present a search for stealth SUSY signatures in final states with two photons from the decays of electroweakinos. The data used in the search correspond to an integrated luminosity of 138\fbinv of proton-proton ($\Pp\Pp$) collisions collected at $\sqrt{s}=13\TeV$ by the CMS experiment in 2016--2018.

\begin{figure}[t]
\centering
    \includegraphics[width=0.45\textwidth]{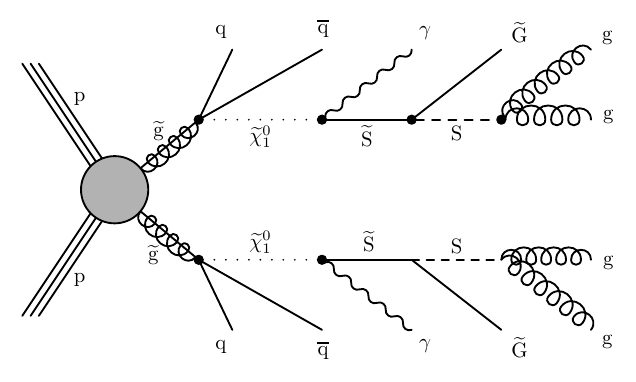}
    \includegraphics[width=0.45\textwidth]{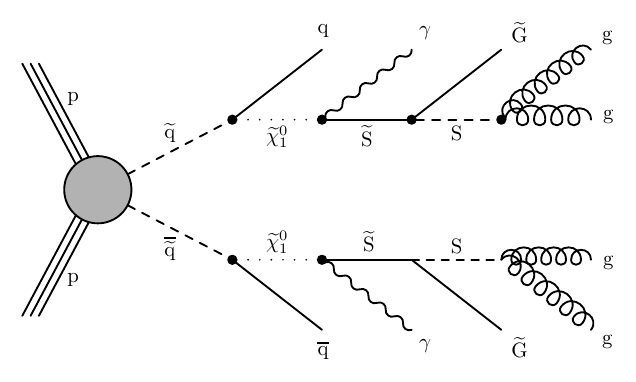}
    \caption{Diagrams of the simplified models considered in this paper. The decay chain starts from the production of either gluino pairs (\cmsLeft) or squark pairs (\cmsRight) and results in a final state consisting of two photons, multiple jets, and low \ptmiss.}
    \label{fig:TNGGStealth}
\end{figure}

We consider models with strongly produced SUSY particles, with pairs either of squarks \sQua or of gluinos \sGlu produced in the primary interaction, as shown in Fig.~\ref{fig:TNGGStealth}. These colored SUSY particles decay to quarks and neutralinos \PSGczDo. Neutralinos further decay to the stealth sector via the emission of photons. Each singlet \PS decays to two SM gluons.

The masses of \PSS~and \PS are parameters of these models, and are fixed to 100 and 90\GeV, respectively, in this study. These values imply a small mass splitting of 10\GeV, as required by the stealth mechanism, and are consistent with stealth benchmark points proposed in Ref.~\cite{Fan:2011yu}. The gravitino \PXXSG is taken to be massless. We consider a two-dimensional scan of models by allowing \gluinoMass~(\squarkMass) to vary in the range $1250<M<2350\GeV$ ($1100<M<2000\GeV$) in increments of 50\GeV. The neutralino mass \neutralinoMass~is varied in changing increments in the range $150\GeV < M < \gluinoMass(\squarkMass) - 100\GeV$, with more points simulated near the boundaries, where the acceptance of the analysis is often sensitive to small changes.

The final state consists of two photons, many jets, and low \ptmiss~carried by the light and soft gravitinos. To obtain an estimate of the SM background, a hybrid approach is used, in which control samples in the data yield a first estimate, which is then adjusted using simulated samples of the major SM processes contributing to the diphoton background.

This paper is organized as follows. In Section~\ref{sec:cmsDetector}, we describe the features of the CMS detector and simulated samples used in this analysis. The trigger criteria and object definitions are given in Section~\ref{sec:trigObjSelect}. The details of the event selection, including the definitions of the search and control regions, are given in Section~\ref{sec:eventSelect}. Section~\ref{sec:backgroundEstimation} discusses the method used to estimate the SM backgrounds in the search, and Section~\ref{sec:systematics} describes the sources of systematic uncertainty. The results of the search, including exclusion limits, are presented in Section~\ref{sec:results}. Section~\ref{sec:summaryConclusion} summarizes our findings. Tabulated results are provided in the HEPData record for this analysis~\cite{hepdata}.

\section{The CMS detector and simulated samples}
\label{sec:cmsDetector}

The central feature of the CMS apparatus is a superconducting solenoid of 6\unit{m} internal diameter, providing a magnetic field of 3.8\unit{T}. Within the solenoid volume are a silicon pixel and strip tracker, a lead tungstate crystal electromagnetic calorimeter (ECAL), and a brass and scintillator hadron calorimeter (HCAL), each composed of a barrel and two endcap sections. Forward calorimeters extend the pseudorapidity ($\eta$) coverage provided by the barrel and endcap detectors. Muons are detected in gas-ionization chambers embedded in the steel flux-return yoke outside the solenoid. A more detailed description of the CMS detector, together with a definition of the coordinate system used and the relevant kinematic variables, can be found in Ref.~\cite{Chatrchyan:2008zzk}.

Monte Carlo (MC) simulation is used to obtain the potential stealth SUSY signal yield. Each of the simplified models introduced in Section~\ref{sec:introduction} is generated with \MGvATNLO~2.4.3 \cite{Alwall:2014hca,Frederix:2018nkq} at leading order (LO), with \PYTHIA~8.230~\cite{Sjostrand:2014zea} used for showering and hadronization. For the simulation of the underlying event, the CP2 tune~\cite{CMS:2019csb} is used. This tune uses the NNPDF3.1LO parton distribution functions (PDFs)~\cite{NNPDF:2017mvq}. The cross sections are calculated with a next-to-next-to-LO (NNLO) approximation up to next-to-next-to-leading logarithmic (NNLL) accuracy~\cite{Fuks:2012qx, Fuks:2013vua, Borschensky:2014cia, Beenakker:2014sma, Beenakker:2016lwe}. The CMS detector response is determined via fast simulation~\cite{Abdullin:2011zz}.

The MC simulation is also used to refine the background estimate based on control samples in data by looking at the sum of three primary contributions to the SM background: diphoton (with both photons reconstructed from the initial hard scattering), $\gamma$+jet (with one photon reconstructed from the initial hard scattering and one from jet fragmentation or from a misidentified jet), and quantum chromodynamics (QCD) multijets (with both photons reconstructed from jet fragmentation or from misidentified jets). All samples are generated separately for 2016, 2017, and 2018 data-taking conditions. In each of the three samples, \PYTHIA~8.230 is used for showering and hadronization. The $\gamma$+jet and QCD multijet samples are generated with \MGvATNLO~2.4.3 at LO and \MGvATNLO~2.6.0 at LO for 2016--2017 and 2018 running conditions, respectively. The 2016 subsamples of both the QCD multijet and $\gamma$+jet samples are generated with the CUETP8M1 tune~\cite{CMS:2015wcf}, which uses the NNPDF3.0LO PDFs~\cite{Ball:2013hta}, while the 2017--2018 subsamples of both the QCD multijet and $\gamma$+jet samples are generated with the CP5 tune~\cite{CMS:2019csb}, which uses the NNPDF3.1LO PDFs. The diphoton samples are generated at LO with \SHERPA~2.1.0~\cite{Sherpa:2019gpd}, \SHERPA~2.2.4, and \SHERPA~2.2.10 for 2016, 2017, and 2018 running conditions, respectively, and use the CT10 PDFs~\cite{Lai:2010vv} from the \textsc{lhapdf} sets~\cite{Buckley:2014ana} for all three years. The CMS detector response is modeled with \GEANTfour~\cite{GEANT4:2002zbu} for all background samples. Additional $\Pp\Pp$ interactions within the same or nearby bunch crossings (pileup) are included in the simulation.

The particle-flow (PF) algorithm~\cite{CMS:2017yfk} aims to reconstruct and identify each particle in an event, with an optimized combination of information from the various elements of the CMS detector. The energies of photons are obtained from the ECAL measurement. The energies of electrons are determined from a combination of the electron momentum at the primary interaction vertex as determined by the tracker, the energy of the corresponding ECAL cluster, and the energy sum of all bremsstrahlung photons spatially compatible with originating from the electron track. The energies of muons are obtained from the curvature of the corresponding track. The energies of charged hadrons are determined from a combination of their momentum measured in the tracker and the matching ECAL and HCAL energy deposits, corrected for the response function of the calorimeters to hadronic showers. Finally, the energies of neutral hadrons are obtained from the corresponding corrected ECAL and HCAL energy deposits.

\section{Trigger and object selection}
\label{sec:trigObjSelect}

Events of interest are selected using a two-tiered trigger system. The first level of the trigger system, composed of custom hardware processors, uses information from the calorimeters and muon detectors to select events at a rate of around 100\unit{kHz} within a fixed latency of about 4\mus~\cite{CMS:2020cmk}. The second level, known as the high-level trigger, consists of a farm of processors running a version of the full event reconstruction software optimized for fast processing, and reduces the event rate to around 1\unit{kHz} before data storage~\cite{CMS:2016ngn}.

As described in Section~\ref{sec:eventSelect}, the analysis uses a diphoton signal selection, a diphoton control selection, and a single-photon control selection. We employ two very similar diphoton triggers, of which one was active in the 2016 data-taking period, and the other in the 2017--2018 data-taking periods. These triggers require leading and subleading photons with thresholds on the transverse momentum \pt~at 30 and 18\GeV, respectively. Both triggers impose a threshold requirement of 55\GeV on the invariant mass of the two photons ($m_{\gamma \gamma}$). The 2016 trigger vetoes events if both photon candidates are associated with charged tracks, to limit contamination from electrons. The 2017--2018 trigger vetoes events if either photon candidate is associated with charged tracks. These diphoton triggers are used to populate the signal selection and the diphoton control selection. In addition, we employ a single-photon trigger, which requires at least one photon with a \pt~threshold of 200\GeV in addition to a loose isolation requirement. The single-photon trigger is used to populate the single-photon control selection.

Jet candidates are reconstructed using PF objects via the anti-\kt clustering algorithm~\cite{Cacciari:2008gp} with a distance parameter of 0.4~\cite{Cacciari:2011ma}. Jet momentum is determined as the vectorial sum of all PF object momenta in the jet, and is found from simulation to be, on average, within 5--10\% of the true momentum over the entire \pt~spectrum and detector acceptance. Pileup can contribute additional tracks and calorimetric energy depositions, increasing the apparent jet momentum. To mitigate this effect, tracks identified to be originating from pileup vertices are discarded and an offset correction is applied to correct for remaining contributions. Jet energy corrections are derived from simulation studies so that the average measured energy of jets becomes identical to that of particle level jets. In-situ measurements of the momentum balance in dijet, $\gamma$+jet, $\PZ$+jet, and multijet events are used to determine any residual differences between the jet energy scale in data and simulation, and appropriate corrections are made~\cite{CMS:2016lmd}. Additional selection criteria are applied to each jet to remove jets potentially dominated by instrumental effects or reconstruction failures. After correction, jet candidates are required to have $\pt>30\GeV$ and to be within the range $\abs{\eta}<2.4$.

Photon candidates are reconstructed from energy clusters in the ECAL. Candidates are required to be reconstructed in the ECAL barrel, with $\abs{\eta}<1.442$. In order to reduce background from misidentification of electrons as photons, candidates are rejected if they have measurements in the pixel detector compatible with the ECAL cluster. This veto eliminates candidates with two or more measurements in the pixel detectors pointing toward the ECAL cluster corresponding to the candidate. For each photon candidate, we compute PF-based isolation variables by summing the \pt~magnitudes of all reconstructed charged and neutral hadrons, and additional photons within an angular cone of $\Delta R \equiv \sqrt{(\Delta \eta)^2 + (\Delta \phi)^2} = 0.3$, $\phi$ being the azimuthal angle in radians. These isolation variables are used to reduce the misidentification of hadrons as photons. Further details on photon identification criteria can be found in Ref.~\cite{CMS:2020uim}. Candidates that meet all kinematic criteria, and tight shower shape and isolation requirements, are classified as nominal photons in this analysis.

The vector \ptvecmiss is defined as the negative vector \pt~sum of all reconstructed PF objects in the event, adjusted for jet energy corrections. The \ptmiss is then the magnitude of this vector. Dedicated filters are applied to remove artificial contributions to \ptmiss from various sources, including the beam halo, noise from detector electronics, poorly reconstructed muons, and poorly calibrated ECAL endcap crystals~\cite{CMS:2019ctu}.

\section{Event selection}
\label{sec:eventSelect}

We define \st as the scalar sum of \ptmiss and the \pt~values of all jets and photons in the event, as selected with the criteria given in Section~\ref{sec:trigObjSelect}. Events with $\st>1200\GeV$, and with two or more reconstructed jets, are considered for the analysis. Based on the number and quality of the selected photon candidates, events are categorized into three selections:

\begin{itemize}
\item The signal selection is constructed from events passing the diphoton trigger preselection requirements. Events are required to have a leading photon candidate with $\pt>35\GeV$ and at least one subleading photon candidate with $\pt>25\GeV$, and the two candidates are required to have $m_{\gamma\gamma} > 90\GeV$. The diphoton invariant mass requirement reduces the background, while retaining at least 98\% signal efficiency for the stealth SUSY models. Both candidates are required to pass the nominal photon criteria described in Section~\ref{sec:trigObjSelect}.
\item The diphoton control selection is constructed from events passing the diphoton trigger preselection. Events are required to have two photon candidates passing the same requirements on the \pt values and $m_{\gamma\gamma}$ as those of the signal selection. For the diphoton control selection, one or both photon candidates are required to fail the tight shower shape or isolation requirements of nominal photons.
\item The single-photon control selection is constructed from events passing the single-photon trigger preselection. The candidate photon is required to have $\pt>200\GeV$ and to pass tight shower shape and isolation requirements.
\end{itemize}

In all events, the jet multiplicity \nJets is defined as the number of selected jet candidates that are more than $\Delta R = 0.4$ away from a selected photon. All three selections are further divided into subregions based on \st and \nJets. Events with exactly two jets compose the shape sideband, which is used to derive the shape of the \st distribution to be used at higher values of \nJets, as described in Section~\ref{sec:backgroundEstimation}. The region $1200<\st<1300\GeV$ is defined as the normalization sideband and is used to obtain the normalization, in each \nJets bin, for the \st shape determined from the shape sideband. In each selection, the search region consists of events with four or more jets and $\st>1300\GeV$, and is divided into three \nJets bins (4, 5, and $\geq$6) and six ranges of \st (1300--1400, 1400--1500, 1500--1700, 1700--2000, 2000--2500, and $>$2500\GeV) for a total of 18 search regions.

\section{Background estimation}
\label{sec:backgroundEstimation}

The SM backgrounds to this search arise mainly from multijet events with two photons produced in the initial scattering. In general, multijet SM backgrounds at high \nJets can be derived using the \st shape invariance method~\cite{CMS:2010oej, CMS:2012grf, CMS:2013ndg, CMS:2017boz}. The key idea is that in SM processes, the fragmentation products of a boosted jet are nearly collinear and therefore have an \st close to that of similar events with fewer reconstructed jets. In other words, the \st distribution of events with low values of \nJets, where the signal contamination is very low for the models studied, can be used to predict the background \st distribution at high values of \nJets. In this analysis, because the selection requires events with two reconstructed photons, deviations from \st scaling could arise from selection efficiency biases. Therefore, we use simulated background samples to estimate first-order corrections to the baseline expectation of \st invariance. The background model used in this analysis can be expressed as follows:
\begin{equation}
  \label{eq:backgroundModel}
  \begin{aligned}
    b(\nJets,\st\,\text{bin}\,i) &= \NEvtsNorm\\
    &\times \fshape\\
    &\times \radjustment,
  \end{aligned}
\end{equation}
where $b(\nJets,\st\,\text{bin}\,i)$ is the expected background in a given $(\nJets,\st)$ bin. Here, \NEvtsNorm is the number of events in a low-\st normalization bin in the selection at a given \nJets, $\fshape$ is a shape template obtained from the data selection at $\nJets=2$, and \radjustment is a correction to the shape template estimated from simulated background samples. These shape and correction factors are described in further detail below.

\begin{figure}[tpb]
\centering
  \includegraphics[width=0.49\textwidth]{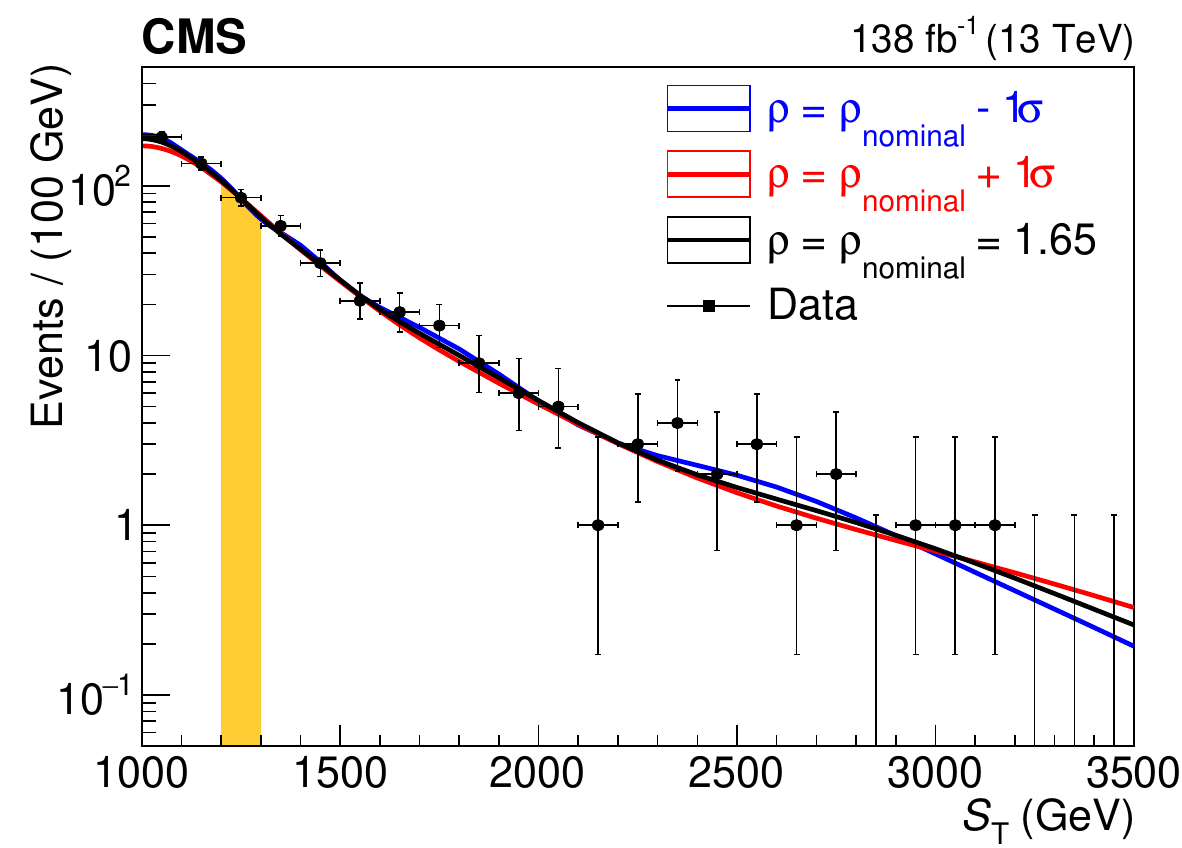}
  \caption{The AGK template of the \st distribution, obtained with various values of $\rho$, from the signal selection in data for $\nJets = 2$. The region shaded in yellow is the normalization sideband.}
  \label{fig:agk}
\end{figure}

The factors $\fshape$ are obtained using the shape in the signal selection of the observed \st distribution for the SM background for $\nJets = 2$. This shape is determined as an adaptive Gaussian kernel (AGK)~\cite{Cranmer:2000du}, which provides an unbinned estimate of the density of the \st distribution. The AGK template depends only on a single parameter $\rho$, which is used to tune the balance between resilience to statistical fluctuations and ability to pick up small-scale features. In order to estimate the optimal value of $\rho$ to use for the \st template, we divide the data set with $\nJets=2$ into three equal sets, and evaluate the negative log likelihood (NLL) for each third of the data set with respect to the kernel obtained from the remaining two sets. In this approach, called 3-fold cross-validation, the value of $\rho$ that minimizes the sum of the NLLs over each third of the data set is taken as the value of $\rho$ to use in the \st~template. The factors $\fshape$ in each \st bin are then obtained as the integral of the AGK template \AGKTemplate~in the given bin, normalized to the integral of the AGK template in the normalization bin. Figure~\ref{fig:agk} shows the data and the AGK template at the nominal value of $\rho$ as well as templates with $\rho$ shifted by one standard deviation ($\pm 1 \sigma$), as described in Section~\ref{sec:systematics}. We note that lower values of $\rho$ allow the kernel to pick up features in the data at smaller scales.

The correction terms \radjustment are found from a sum of three simulated backgrounds, those corresponding to diphoton, $\gamma+$jet, and QCD multijet processes. We first obtain the nominal shape template from the combined background at $\nJets = 2$. At $\nJets \geq 4$, a modified version of the nominal shape template is fit to the \st distribution in the combined background. More specifically, the template at $\nJets \geq 4$ is taken to be the nominal shape template multiplied by a simple linear adjustment of the form $[A + m(\frac{\st}{\st^{\text{norm}}} - 1)]$ (where $\st^{\text{norm}} = 1250\GeV$ is the center of the normalization bin). The $A$ and $m$ values for each \nJets bin are the parameters of the fit, and their best fit values can be found in Table~\ref{tab:bestfits}. Finally, the factors \radjustment are evaluated as the ratio of the corrected to the uncorrected prediction in a given \st bin. We note that the factors \radjustment encode the deviation from \st scaling in the combined simulated background.

In each $(\nJets,\st)$ bin, the signal prediction is obtained directly from a signal MC sample. Because the method used to obtain the background is based on control samples in the data, the background estimate is susceptible to potential signal contamination. To correct for this, we evaluate the background contamination and subtract it from the signal prediction in each $(\nJets, \st)$ bin. In addition, we limit inference of upper limits, as detailed in Section~\ref{sec:results}, to those regions of the $(\gluinoMass(\squarkMass), \neutralinoMass)$ parameter space in which the potential signal contamination in the sidebands is below 10\%, by eliminating the region with $\gluinoMass (\squarkMass) - \neutralinoMass < 100\GeV$. Furthermore, the signal efficiency is quite small in the region of the phase space with very low \neutralinoMass~(at high values of \nJets, the signal efficiency goes from ${\approx}$7\% at low \neutralinoMass to ${\approx}$30\% at higher \neutralinoMass), because in such events the neutralino is highly Lorentz-boosted, and its decay results in collimated jets and photons. Therefore, for inference, we do not consider the phase space below $\neutralinoMass = 125 (200)\GeV$ for gluino (squark) samples.

\begin{table}[tbh]
\centering
\topcaption{Best fit values of $A$ and $m$ with statistical $\pm 1\sigma$ errors, obtained from the simulated background.}
\begin{scotch}{lll}
  Best fit values & $A$ & $m$ \\ \hline
  \nJets = 4 & 1.04 $\pm$ 0.03 & 0.75 $\pm$ 0.09 \\
  \nJets = 5 & 0.99 $\pm$ 0.04 & 1.30 $\pm$ 0.15 \\
  \nJets $\geq$ 6 & 1.11 $\pm$ 0.06 & 2.42 $\pm$ 0.23 \\
\end{scotch}
\label{tab:bestfits}
\end{table}

\section{Systematic uncertainties}
\label{sec:systematics}

The background prediction is obtained from Eq.~(\ref{eq:backgroundModel}), each factor of which has potential uncertainties that affect the background yield. The signal prediction, being estimated from simulation, also has several potential uncertainties that affect the signal yield. In the following paragraphs, we describe the methods used to evaluate the effect of each source of uncertainty on the overall yield.

First, we describe the uncertainties affecting background yield. In Eq.~(\ref{eq:backgroundModel}), there exists a systematic uncertainty in the prediction arising from the statistical uncertainty in \NEvtsNorm. We use a 68\% Poisson confidence interval around the observed value of \NEvtsNorm to estimate the effect of this uncertainty, which we find to be in the range 15--22\%, depending on the \nJets bin.

There are two sources of uncertainty in the shape factors $\fshape$ in Eq.~(\ref{eq:backgroundModel}). First, because of a limited number of events in the data at $\nJets=2$, the template shape estimated as an AGK from the data is subject to statistical fluctuations. To evaluate this effect, we use the nominal template as a probability distribution function to generate simulated pseudo-data sets, where the number of events in each search region is chosen independently. The resulting distribution of $\fshape$ in each \st bin is then used to determine the relative uncertainty in the nominal prediction. This uncertainty is in the range 10--25\%, depending on the \st bin. Second, there is an uncertainty introduced by the choice of the parameter $\rho$ of the template. This uncertainty is estimated by varying the nominal value of $\rho$ by its $\pm 1\sigma$ fluctuations obtained from the NLL curve of the 3-fold cross-validation, and finding the relative change in $\fshape$. This uncertainty is less than 3\% for all bins.

There are three sources of uncertainty in the corrections \radjustment in Eq.~(\ref{eq:backgroundModel}). First, the best fit corrections that enter Eq.~(\ref{eq:backgroundModel}) are subject to statistical fluctuations because they are estimated from a statistically limited simulated background sample. This uncertainty is evaluated by diagonalizing the covariance matrix of the fits to the \st distribution in each \nJets bin. The fluctuations of \radjustment for each \nJets bin, estimated as variations of the best fit parameters consistent with the covariance matrix, are used as a measure of this uncertainty. This procedure yields uncertainty estimates of less than 7\% in all bins. Next, because these corrections are estimated from a simulated sample, we account for the potential mismodeling in the simulation both as a function of \st and as a function of \nJets, each of which is treated as a separate source of uncertainty. In order to estimate the uncertainty in \radjustment due to mismodeling as a function of \nJets, we vary the cross section of each of the three backgrounds to the combination up and down by a factor of 2, thereby varying the relative composition of the simulated background, and the variation in \radjustment is then used to determine the uncertainty. This uncertainty in \radjustment leads to an uncertainty in the background prediction of less than 8\% in all bins. Next, in order to estimate the uncertainty in \radjustment due to mismodeling as a function of \st, we consider two event selections orthogonal to the diphoton selections used in this analysis: a selection of events with exactly one selected photon, and a selection of events that pass the trigger but do not pass the event selection. In both selections, we obtain the ratio of \st distributions at high values of \nJets to the \st~distribution at $\nJets=2$ both for the data and for the simulated background. A ratio of the 2-to-$n$ correction in data, \radjustment, to this correction in the simulated background, is a measure of the nonclosure of this procedure in both orthogonal selections, and is used to estimate the uncertainty in the background prediction due to mismodeling as a function of \st. This leads to a flat 10\% uncertainty in all signal bins.

The uncertainties in the signal prediction are typically much smaller than the uncertainties in the background. There is a systematic uncertainty related to the finite size of the simulated signal samples (1--5\%, depending on the signal point). Next, there are systematic uncertainties due to the uncertainty in the knowledge of the jet energy scale (about 10\%), due to a potential mismeasurement of the integrated luminosity (1.6\%)~\cite{CMS:2021xjt,CMSPAS:lumi2017,CMSPAS:lumi2018}, and due to uncertainties in the trigger efficiency (2--10\%). Other uncertainties, such as in the photon reconstruction efficiency and in the value of \ptmiss due to the jet energy resolution and unclustered energy in the event, are negligible.

\section{Results}
\label{sec:results}

\begin{figure}[htbp]
  \centering
  \includegraphics[width=0.49\textwidth]{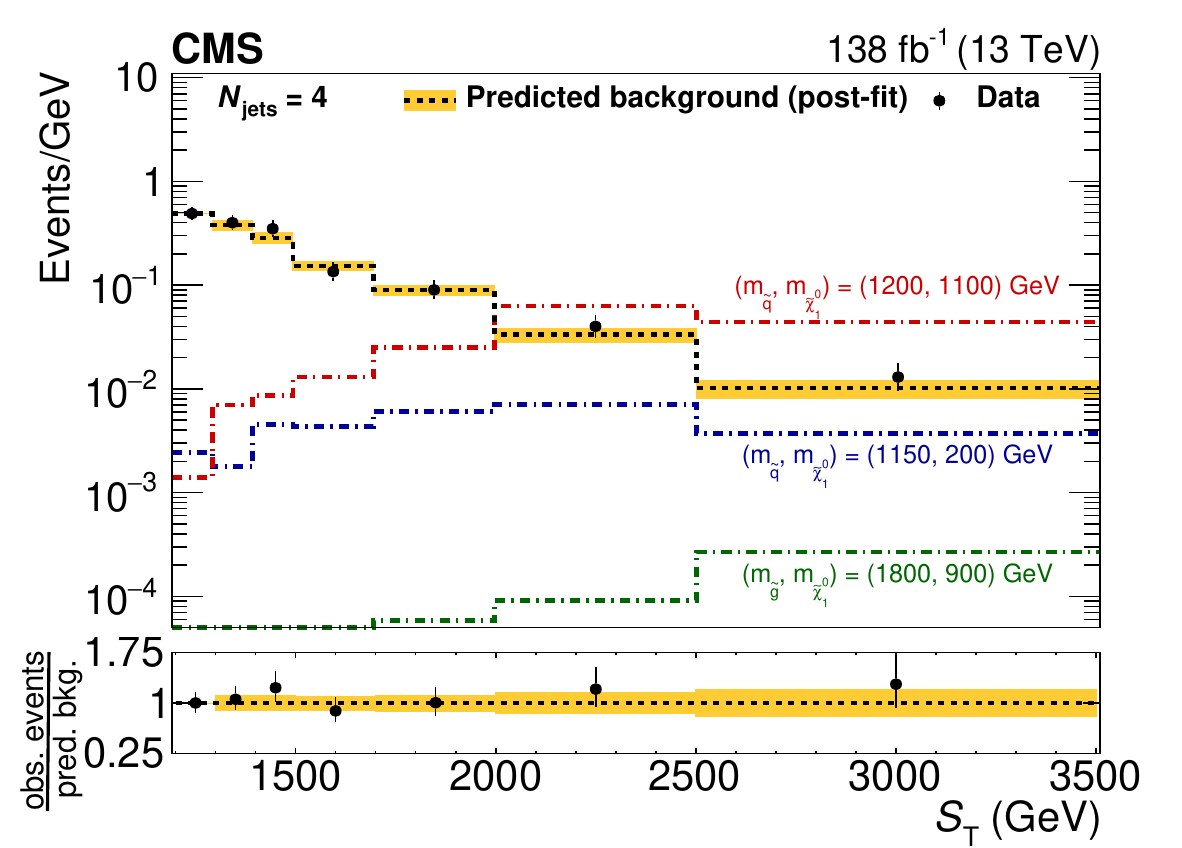}\\
  \includegraphics[width=0.49\textwidth]{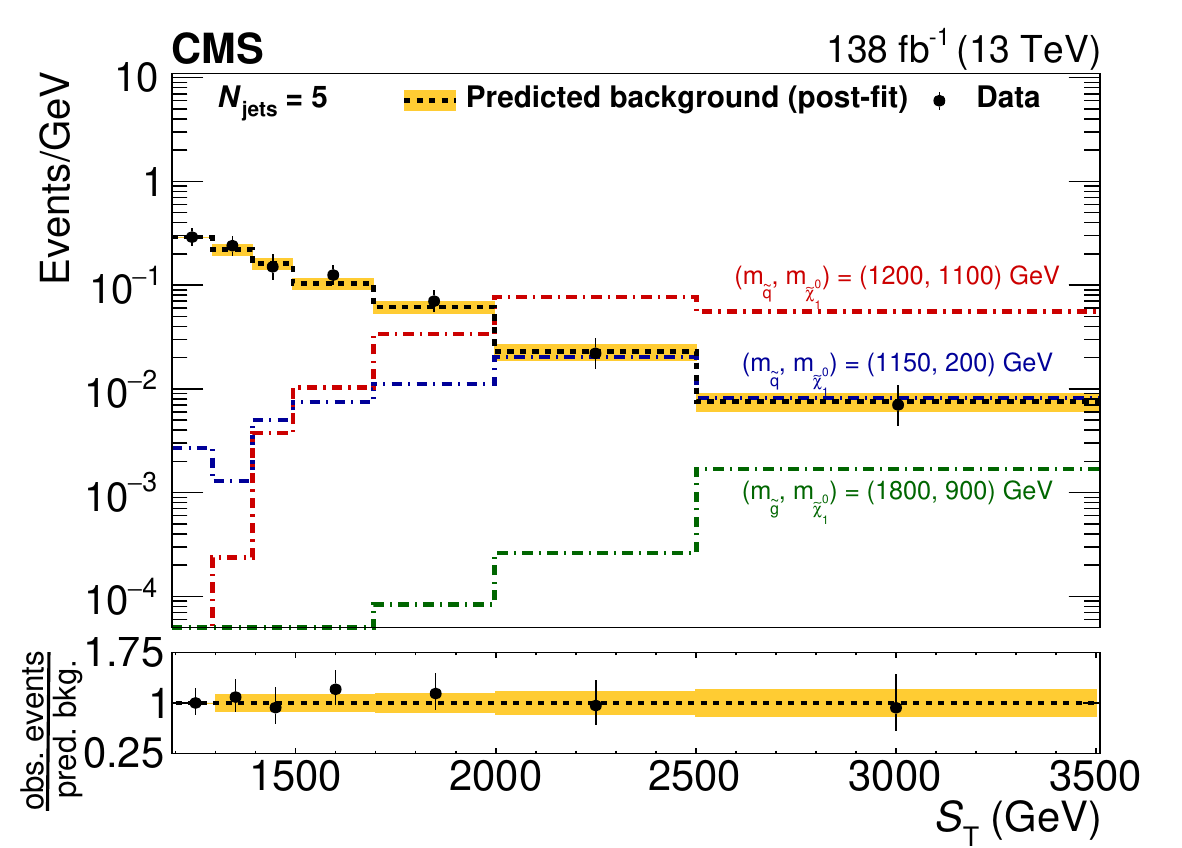}\\
  \includegraphics[width=0.49\textwidth]{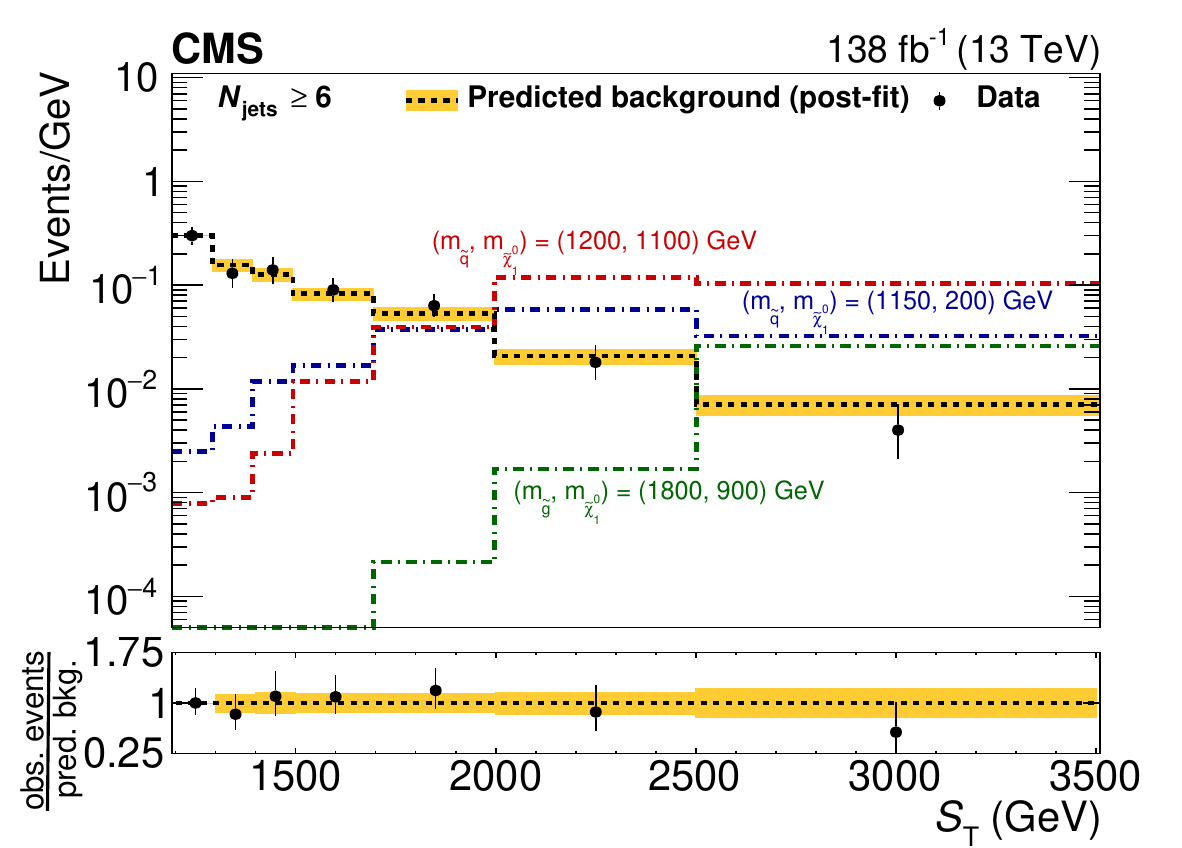}
  \caption{Comparison of the measured \st distribution with the post-fit background prediction in each of the search regions of the analysis for \nJets = 4 (upper), 5 (middle), and $\geq$6 (lower). The post-fit uncertainties on the background prediction are represented by the yellow band around the central prediction. Signal yields at a signal strength of 1 are also overlaid for some representative points in the $(\gluinoMass(\squarkMass), \neutralinoMass)$ parameter space.}
  \label{fig:fullChainResult}
\end{figure}

The yield in a given $(\nJets,\st)$ bin is modeled as $b + \mu s$, where $b$ is the background estimate, $s$ is the signal expectation, and $\mu$ is the signal strength. Both $b$ and $s$ are functions of nuisance parameters that model the various uncertainties described in Section~\ref{sec:systematics}. A comparison of the measured \st distribution and the post-fit background prediction (\ie, the background prediction obtained by setting all nuisance parameters to those specific values that maximize the likelihood function at zero signal strength), with post-fit uncertainties, is shown in Fig.~\ref{fig:fullChainResult} for each \nJets bin. The data reveal no indication of any deviation from the background prediction.

The results are interpreted as 95\% confidence level (\CL) upper limits on the gluino and squark pair production cross sections of the simplified stealth SUSY models described in Section~\ref{sec:introduction}. For the statistical inference of these upper limits, we use the \CLs criterion~\cite{Junk:1999kv, Read:2002hq}, which is based on a log-likelihood ratio test statistic, modified for upper limits~\cite{CMSNote:combine}, that compares the SM-only hypothesis to the hypothesis that there is an additional contribution from the signal. An asymptotic approximation~\cite{Cowan:2010js} for the test statistic is used to extract the observed limits. Systematic uncertainties are incorporated into the test statistic as nuisance parameters. The background prediction, signal expectation, and observed number of events in each search region are combined into a single statistical interpretation, and evaluated as a multichannel counting experiment.

\begin{figure}[bht]
  \centering
  \includegraphics[width=0.47\textwidth]{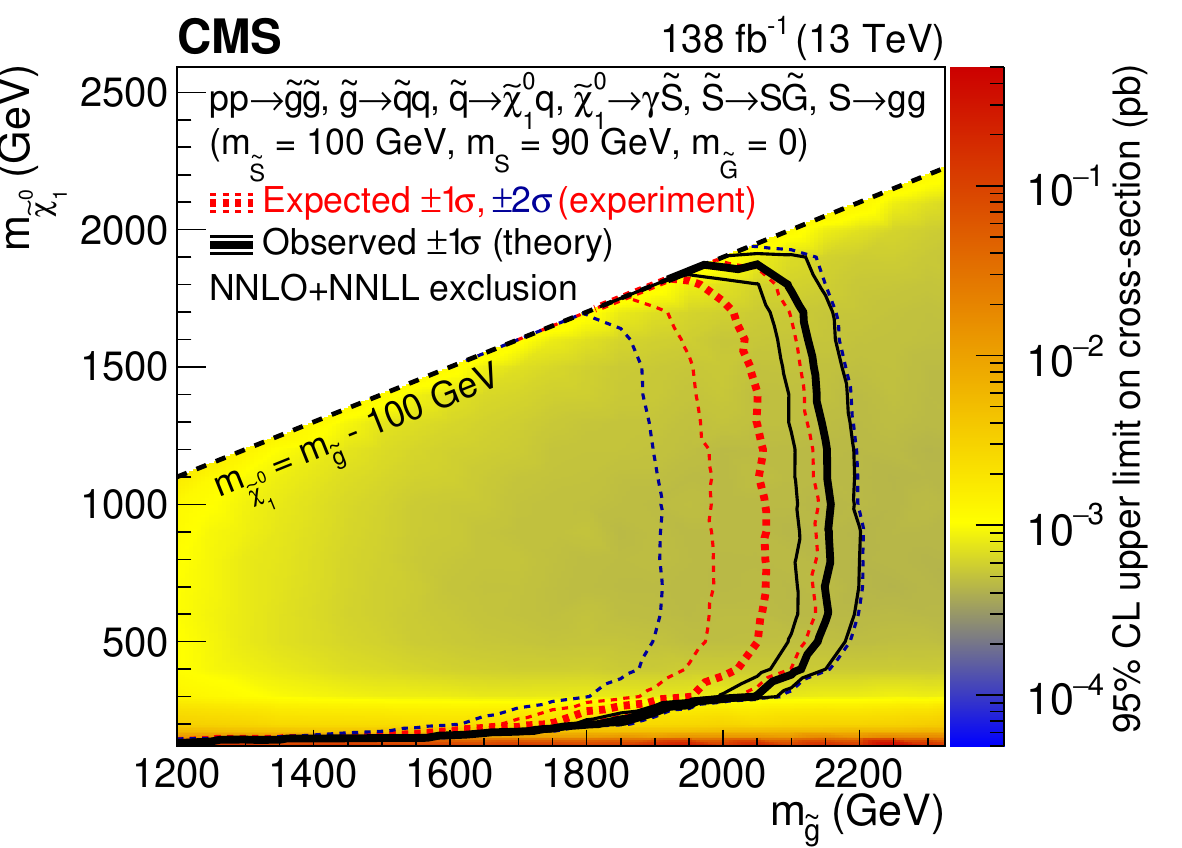}
  \includegraphics[width=0.47\textwidth]{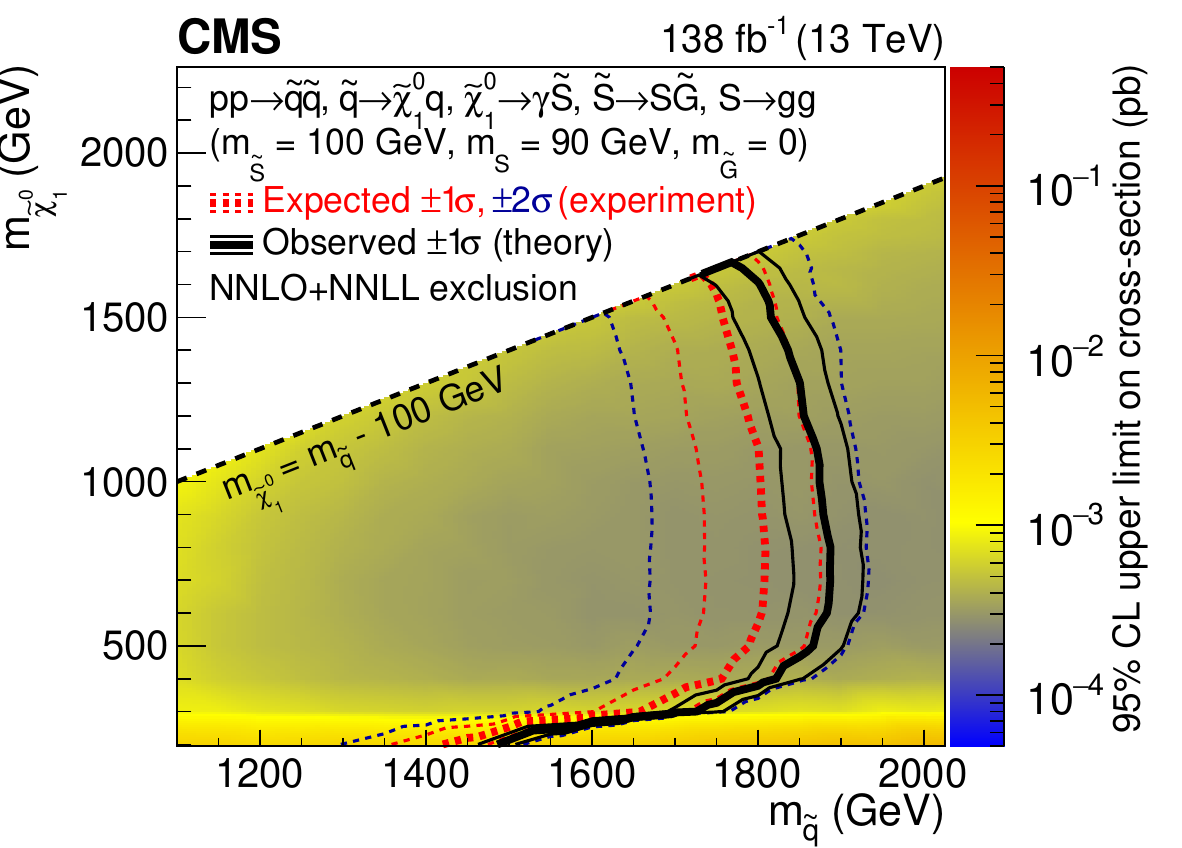}
\caption{The 95\% \CL upper limits on the gluino pair production cross section (\cmsLeft) and squark pair production cross section (\cmsRight) as functions of \gluinoMass (\squarkMass) and \neutralinoMass. The contours show the observed (solid line) and expected (dashed line) 95\% \CL exclusions with their one standard deviation uncertainties. Two standard deviation uncertainties are also shown for the expected 95\% \CL exclusions.}
  \label{fig:fullLimitPlot}
\end{figure}

The upper limit on the gluino (squark) pair production cross section is shown in Fig.~\ref{fig:fullLimitPlot} as a function of \gluinoMass (\squarkMass) and \neutralinoMass, for the simplified stealth models used in this analysis. The production cross sections of the model points are determined exclusively by the value of \gluinoMass (\squarkMass). For most \PSGczDo masses, we exclude \gluinoMass up to 2150 or \squarkMass up to 1850\GeV at 95\% \CL. Despite a relaxation of assumptions in the background modeling as compared to previously published results, we achieve a ${\approx}$70\% improvement in the reach of the exclusion contour in the $(\squarkMass, \neutralinoMass)$ parameter space, where \neutralinoMass is the neutralino mass.

In the $(\nJets=6, \st > 2500\GeV)$ bin, we observe fewer events than predicted by the background model, making the observed limits stronger than the expected limits for most values of \gluinoMass (\squarkMass) and \neutralinoMass. In regions of the parameter space at low \neutralinoMass, the neutralino becomes boosted, and the resulting decay products become collimated. This has the effect of causing the photon to fail the isolation criteria, and results in a weakening of the exclusion limits. In regions of the parameter space where \neutralinoMass is very close to \gluinoMass (\squarkMass), the jets produced from the gluino decay become soft. This results in a reduction in the overall number of reconstructed jets in the event, and also causes a reduction in the excluded \gluinoMass.

\section{Summary}
\label{sec:summaryConclusion}

A search for stealth supersymmetry in events with two photons and at least four jets is presented. No threshold requirement on the missing transverse momentum is imposed. The measurement is based on a data sample corresponding to an integrated luminosity of 138\fbinv of proton-proton collisions at $\sqrt{s} = 13\TeV$ recorded by the CMS experiment in 2016--2018. The analysis is performed in bins of \st, which is the sum of transverse momenta for all reconstructed particles in the event, as well as jet multiplicity \nJets. A background model based on control samples in data is employed to estimate the standard model (SM) background in the data. The data are found to be consistent with the SM background prediction with no evidence of significant signal contribution. The results of the search are interpreted as 95\% confidence level upper limits on the gluino and squark production cross sections in the context of simplified models of stealth supersymmetry. In simplified stealth models with gluino (squark) pair production, we exclude gluino masses \gluinoMass (squark masses \squarkMass) up to 2150 (1850)\GeV. These are the most stringent limits to date on these models.

\begin{acknowledgments}
  We congratulate our colleagues in the CERN accelerator departments for the excellent performance of the LHC and thank the technical and administrative staffs at CERN and at other CMS institutes for their contributions to the success of the CMS effort. In addition, we gratefully acknowledge the computing centers and personnel of the Worldwide LHC Computing Grid and other centers for delivering so effectively the computing infrastructure essential to our analyses. Finally, we acknowledge the enduring support for the construction and operation of the LHC, the CMS detector, and the supporting computing infrastructure provided by the following funding agencies: SC (Armenia), BMBWF and FWF (Austria); FNRS and FWO (Belgium); CNPq, CAPES, FAPERJ, FAPERGS, and FAPESP (Brazil); MES and BNSF (Bulgaria); CERN; CAS, MoST, and NSFC (China); MINCIENCIAS (Colombia); MSES and CSF (Croatia); RIF (Cyprus); SENESCYT (Ecuador); MoER, ERC PUT and ERDF (Estonia); Academy of Finland, MEC, and HIP (Finland); CEA and CNRS/IN2P3 (France); SRNSF (Georgia); BMBF, DFG, and HGF (Germany); GSRI (Greece); NKFIH (Hungary); DAE and DST (India); IPM (Iran); SFI (Ireland); INFN (Italy); MSIP and NRF (Republic of Korea); MES (Latvia); LAS (Lithuania); MOE and UM (Malaysia); BUAP, CINVESTAV, CONACYT, LNS, SEP, and UASLP-FAI (Mexico); MOS (Montenegro); MBIE (New Zealand); PAEC (Pakistan); MES and NSC (Poland); FCT (Portugal); MESTD (Serbia); MCIN/AEI and PCTI (Spain); MOSTR (Sri Lanka); Swiss Funding Agencies (Switzerland); MST (Taipei); MHESI and NSTDA (Thailand); TUBITAK and TENMAK (Turkey); NASU (Ukraine); STFC (United Kingdom); DOE and NSF (USA).
    
  \hyphenation{Rachada-pisek} Individuals have received support from the Marie-Curie program and the European Research Council and Horizon 2020 Grant, contract Nos.\ 675440, 724704, 752730, 758316, 765710, 824093, and COST Action CA16108 (European Union); the Leventis Foundation; the Alfred P.\ Sloan Foundation; the Alexander von Humboldt Foundation; the Science Committee, project no. 22rl-037 (Armenia); the Belgian Federal Science Policy Office; the Fonds pour la Formation \`a la Recherche dans l'Industrie et dans l'Agriculture (FRIA-Belgium); the Agentschap voor Innovatie door Wetenschap en Technologie (IWT-Belgium); the F.R.S.-FNRS and FWO (Belgium) under the ``Excellence of Science -- EOS" -- be.h project n.\ 30820817; the Beijing Municipal Science \& Technology Commission, No. Z191100007219010 and Fundamental Research Funds for the Central Universities (China); the Ministry of Education, Youth and Sports (MEYS) of the Czech Republic; the Shota Rustaveli National Science Foundation, grant FR-22-985 (Georgia); the Deutsche Forschungsgemeinschaft (DFG), under Germany's Excellence Strategy -- EXC 2121 ``Quantum Universe" -- 390833306, and under project number 400140256 - GRK2497; the Hellenic Foundation for Research and Innovation (HFRI), Project Number 2288 (Greece); the Hungarian Academy of Sciences, the New National Excellence Program - \'UNKP, the NKFIH research grants K 124845, K 124850, K 128713, K 128786, K 129058, K 131991, K 133046, K 138136, K 143460, K 143477, 2020-2.2.1-ED-2021-00181, and TKP2021-NKTA-64 (Hungary); the Council of Science and Industrial Research, India; ICSC -- National Research Center for High Performance Computing, Big Data and Quantum Computing, funded by the EU NexGeneration program (Italy); the Latvian Council of Science; the Ministry of Education and Science, project no. 2022/WK/14, and the National Science Center, contracts Opus 2021/41/B/ST2/01369 and 2021/43/B/ST2/01552 (Poland); the Funda\c{c}\~ao para a Ci\^encia e a Tecnologia, grant CEECIND/01334/2018 (Portugal); the National Priorities Research Program by Qatar National Research Fund; MCIN/AEI/10.13039/501100011033, ERDF ``a way of making Europe", and the Programa Estatal de Fomento de la Investigaci{\'o}n Cient{\'i}fica y T{\'e}cnica de Excelencia Mar\'{\i}a de Maeztu, grant MDM-2017-0765 and Programa Severo Ochoa del Principado de Asturias (Spain); the Chulalongkorn Academic into Its 2nd Century Project Advancement Project, and the National Science, Research and Innovation Fund via the Program Management Unit for Human Resources \& Institutional Development, Research and Innovation, grant B05F650021 (Thailand); the Kavli Foundation; the Nvidia Corporation; the SuperMicro Corporation; the Welch Foundation, contract C-1845; and the Weston Havens Foundation (USA).
\end{acknowledgments}

\bibliography{auto_generated}
\cleardoublepage \appendix\section{The CMS Collaboration \label{app:collab}}\begin{sloppypar}\hyphenpenalty=5000\widowpenalty=500\clubpenalty=5000
\cmsinstitute{Yerevan Physics Institute, Yerevan, Armenia}
{\tolerance=6000
A.~Hayrapetyan, A.~Tumasyan\cmsAuthorMark{1}\cmsorcid{0009-0000-0684-6742}
\par}
\cmsinstitute{Institut f\"{u}r Hochenergiephysik, Vienna, Austria}
{\tolerance=6000
W.~Adam\cmsorcid{0000-0001-9099-4341}, J.W.~Andrejkovic, T.~Bergauer\cmsorcid{0000-0002-5786-0293}, S.~Chatterjee\cmsorcid{0000-0003-2660-0349}, K.~Damanakis\cmsorcid{0000-0001-5389-2872}, M.~Dragicevic\cmsorcid{0000-0003-1967-6783}, A.~Escalante~Del~Valle\cmsorcid{0000-0002-9702-6359}, P.S.~Hussain\cmsorcid{0000-0002-4825-5278}, M.~Jeitler\cmsAuthorMark{2}\cmsorcid{0000-0002-5141-9560}, N.~Krammer\cmsorcid{0000-0002-0548-0985}, D.~Liko\cmsorcid{0000-0002-3380-473X}, I.~Mikulec\cmsorcid{0000-0003-0385-2746}, J.~Schieck\cmsAuthorMark{2}\cmsorcid{0000-0002-1058-8093}, R.~Sch\"{o}fbeck\cmsorcid{0000-0002-2332-8784}, D.~Schwarz\cmsorcid{0000-0002-3821-7331}, M.~Sonawane\cmsorcid{0000-0003-0510-7010}, S.~Templ\cmsorcid{0000-0003-3137-5692}, W.~Waltenberger\cmsorcid{0000-0002-6215-7228}, C.-E.~Wulz\cmsAuthorMark{2}\cmsorcid{0000-0001-9226-5812}
% ...
\par}
\vskip\cmsinstskip
\dag:~Deceased\\
$^{1}$Also at Yerevan State University, Yerevan, Armenia\\
$^{2}$Also at TU Wien, Vienna, Austria\\
% ...
\end{sloppypar}
\end{document}